\documentclass[letterpaper,english,reprint, aps, ]{revtex4-2}
\usepackage[T1]{fontenc}
\usepackage[latin9]{inputenc}
\usepackage{amsmath}
\usepackage{amssymb}
\usepackage{graphicx}
\usepackage{xcolor}



\usepackage{babel}
\begin{document}
\global\long\def\ket#1{\left| #1 \right\rangle }%
 
\global\long\def\mcH{{\mathcal{H}}}%
 
\global\long\def\lrp#1{\left( #1 \right)}%
 
\global\long\def\lrb#1{\left[ #1 \right]}%
 
\global\long\def\lrc#1{\left\{  #1 \right\}  }%
\global\long\def\hatU{\hat{U}}%
 
\global\long\def\cohU{\hat{U}^{\dagger}}%

\global\long\def\bra#1{\left\langle #1\right|}%
\global\long\def\avg#1{\left\langle #1 \right\rangle }%
 
\global\long\def\coa{\hat{a}^{\dagger}}%
 
\global\long\def\aoa{\hat{a}}%
 
\global\long\def\cob{\hat{b}^{\dagger}}%
 
\global\long\def\aob{\hat{b}}%
 
\global\long\def\coc{\hat{c}^{\dagger}}%
 
\global\long\def\aoc{\hat{c}}%
 
\global\long\def\cod{\hat{d}^{\dagger}}%
 
\global\long\def\aod{\hat{d}}%
 
\global\long\def\hatD{\hat{D}}%
 
\global\long\def\hatH{\hat{H}}%
\global\long\def\bfk{{\bf k}}%
\global\long\def\aoA{\hat{A}}%
 
\global\long\def\aoB{\hat{B}}%
\global\long\def\hatQ{\hat{Q}}%
 
\global\long\def\bfM{{\bf M}}%
\global\long\def\bfalp{{\bf \mathbf{\alpha}}}%
\global\long\def\tr{\mathrm{Tr}}%
\global\long\def\half{\mathrm{\frac{1}{2}}}%
 
\global\long\def\bfI{{\bf I}}%

\global\long\def\mcA{{\mathcal{A}}}%
 
\global\long\def\mcB{{\mathcal{B}}}%
 
\global\long\def\mcC{{\mathcal{C}}}%
 
\global\long\def\mcD{{\mathcal{D}}}%
 
\global\long\def\mcE{{\mathcal{E}}}%
 
\global\long\def\mcF{{\mathcal{F}}}%
 
\global\long\def\mcG{{\mathcal{G}}}%
 
\global\long\def\mcH{{\mathcal{H}}}%
 
\global\long\def\mcI{{\mathcal{I}}}%
 
\global\long\def\mcJ{{\mathcal{J}}}%
 
\global\long\def\mcK{{\mathcal{K}}}%
 
\global\long\def\mcL{{\mathcal{L}}}%
 
\global\long\def\mcM{{\mathcal{M}}}%
 
\global\long\def\mcN{\mathcal{N}}%
 
\global\long\def\mcO{{\mathcal{O}}}%
 
\global\long\def\mcP{{\mathcal{P}}}%
 
\global\long\def\mcQ{{\mathcal{Q}}}%
 
\global\long\def\mcR{{\mathcal{R}}}%
 
\global\long\def\mcS{{\mathcal{S}}}%
 
\global\long\def\mcT{\mathcal{T}}%
 
\global\long\def\mcU{{\mathcal{U}}}%
 
\global\long\def\mcV{{\mathcal{V}}}%
 
\global\long\def\mcW{{\mathcal{W}}}%
 
\global\long\def\mcX{{\mathcal{X}}}%
 
\global\long\def\mcY{{\mathcal{Y}}}%
 
\global\long\def\mcZ{{\mathcal{Z}}}%

\global\long\def\bbA{{\mathbb{A}}}%
 
\global\long\def\bbB{{\mathbb{B}}}%
 
\global\long\def\bbC{{\mathbb{C}}}%
 
\global\long\def\bbD{{\mathbb{D}}}%
 
\global\long\def\bbE{{\mathbb{E}}}%
 
\global\long\def\bbF{{\mathbb{F}}}%
 
\global\long\def\bbG{{\mathbb{G}}}%
 
\global\long\def\bbH{{\mathbb{H}}}%
 
\global\long\def\bbI{{\mathbb{I}}}%
 
\global\long\def\bbJ{{\mathbb{J}}}%
 
\global\long\def\bbK{{\mathbb{K}}}%
 
\global\long\def\bbL{{\mathbb{L}}}%
 
\global\long\def\bbM{{\mathbb{M}}}%
 
\global\long\def\bbN{{\mathbb{N}}}%
 
\global\long\def\bbO{{\mathbb{O}}}%
 
\global\long\def\bbP{{\mathbb{P}}}%
 
\global\long\def\bbQ{{\mathbb{Q}}}%
 
\global\long\def\bbR{{\mathbb{R}}}%
 
\global\long\def\bbS{{\mathbb{S}}}%
 
\global\long\def\bbT{\mathbb{T}}%
 
\global\long\def\bbU{{\mathbb{U}}}%
 
\global\long\def\bbV{{\mathbb{V}}}%
 
\global\long\def\bbW{{\mathbb{W}}}%
 
\global\long\def\bbX{{\mathbb{X}}}%
 
\global\long\def\bbY{{\mathbb{Y}}}%
 
\global\long\def\bbZ{{\mathbb{Z}}}%

\global\long\def\mfa{{\mathfrak{a}}}%
 
\global\long\def\mfb{{\mathfrak{b}}}%
 
\global\long\def\mfc{{\mathfrak{c}}}%
 
\global\long\def\mfd{{\mathfrak{d}}}%
 
\global\long\def\mfe{{\mathfrak{e}}}%
 
\global\long\def\mff{{\mathfrak{f}}}%
 
\global\long\def\mfg{{\mathfrak{g}}}%
 
\global\long\def\mfh{{\mathfrak{h}}}%
 
\global\long\def\mfi{{\mathfrak{i}}}%
 
\global\long\def\mfj{{\mathfrak{j}}}%
 
\global\long\def\mfk{{\mathfrak{k}}}%
 
\global\long\def\mfl{{\mathfrak{l}}}%
 
\global\long\def\mfm{{\mathfrak{m}}}%
 
\global\long\def\mfn{{\mathfrak{n}}}%
 
\global\long\def\mfo{{\mathfrak{o}}}%
 
\global\long\def\mfp{{\mathfrak{p}}}%
 
\global\long\def\mfq{{\mathfrak{q}}}%
 
\global\long\def\mfr{{\mathfrak{r}}}%
 
\global\long\def\mfs{{\mathfrak{s}}}%
 
\global\long\def\mft{{\mathfrak{t}}}%
 
\global\long\def\mfu{{\mathfrak{u}}}%
 
\global\long\def\mfv{{\mathfrak{v}}}%
 
\global\long\def\mfw{{\mathfrak{w}}}%
 
\global\long\def\mfx{{\mathfrak{x}}}%
 
\global\long\def\mfy{{\mathfrak{y}}}%
 
\global\long\def\mfz{{\mathfrak{z}}}%

\global\long\def\mfA{{\mathfrak{A}}}%
 
\global\long\def\mfB{{\mathfrak{B}}}%
 
\global\long\def\mfC{{\mathfrak{C}}}%
 
\global\long\def\mfD{{\mathfrak{D}}}%
 
\global\long\def\mfE{{\mathfrak{E}}}%
 
\global\long\def\mfF{{\mathfrak{F}}}%
 
\global\long\def\mfG{{\mathfrak{G}}}%
 
\global\long\def\mfH{{\mathfrak{H}}}%
 
\global\long\def\mfI{{\mathfrak{I}}}%
 
\global\long\def\mfJ{{\mathfrak{J}}}%
 
\global\long\def\mfK{{\mathfrak{K}}}%
 
\global\long\def\mfL{{\mathfrak{L}}}%
 
\global\long\def\mfM{{\mathfrak{M}}}%
 
\global\long\def\mfN{{\mathfrak{N}}}%
 
\global\long\def\mfO{{\mathfrak{O}}}%
 
\global\long\def\mfP{{\mathfrak{P}}}%
 
\global\long\def\mfQ{{\mathfrak{Q}}}%
 
\global\long\def\mfR{{\mathfrak{R}}}%
 
\global\long\def\mfS{{\mathfrak{S}}}%
 
\global\long\def\mfT{{\mathfrak{T}}}%
 
\global\long\def\mfU{{\mathfrak{U}}}%
 
\global\long\def\mfV{{\mathfrak{V}}}%
 
\global\long\def\mfW{{\mathfrak{W}}}%
 
\global\long\def\mfX{{\mathfrak{X}}}%
 
\global\long\def\mfY{{\mathfrak{Y}}}%
 
\global\long\def\mfZ{{\mathfrak{Z}}}%

\global\long\def\mrA{{\mathrm{A}}}%
 
\global\long\def\mrB{{\mathrm{B}}}%
 
\global\long\def\mrC{{\mathrm{C}}}%
 
\global\long\def\mrD{{\mathrm{D}}}%
 
\global\long\def\mrE{{\mathrm{E}}}%
 
\global\long\def\mrF{{\mathrm{F}}}%
 
\global\long\def\mrG{{\mathrm{G}}}%
 
\global\long\def\mrH{{\mathrm{H}}}%
 
\global\long\def\mrI{{\mathrm{I}}}%
 
\global\long\def\mrJ{{\mathrm{J}}}%
 
\global\long\def\mrK{{\mathrm{K}}}%
 
\global\long\def\mrL{{\mathrm{L}}}%
 
\global\long\def\mrM{{\mathrm{M}}}%
 
\global\long\def\mrN{{\mathrm{N}}}%
 
\global\long\def\mrO{{\mathrm{O}}}%
 
\global\long\def\mrP{{\mathrm{P}}}%
 
\global\long\def\mrQ{{\mathrm{Q}}}%
 
\global\long\def\mrR{{\mathrm{R}}}%
 
\global\long\def\mrS{{\mathrm{S}}}%
 
\global\long\def\mrT{{\mathrm{T}}}%
 
\global\long\def\mrU{{\mathrm{U}}}%
 
\global\long\def\mrV{{\mathrm{V}}}%
 
\global\long\def\mrW{{\mathrm{W}}}%
 
\global\long\def\mrX{{\mathrm{X}}}%
 
\global\long\def\mrY{{\mathrm{Y}}}%
 
\global\long\def\mrZ{{\mathrm{Z}}}%

\global\long\def\msG{\mathscr{G}}%

\preprint{APS/123-QED}
\title{Work statistics across a quantum critical surface}
\author{Fan Zhang}
\affiliation{School of Physics, Peking University, Beijing, 100871, China}
\author{H. T. Quan}
\email{Corresponding author: htquan@pku.edu.cn}

\affiliation{School of Physics, Peking University, Beijing, 100871, China }
\affiliation{Collaborative Innovation Center of Quantum Matter, Beijing 100871,
China}
\affiliation{Frontiers Science Center for Nano-optoelectronics, Peking University,
Beijing, 100871, China}
\date{\today}
\begin{abstract}
We study the universality of work statistics of a system quenched
through a quantum critical surface. By using the adiabatic perturbation
theory, we obtain the general scaling behavior for all cumulants of
work. These results extend the studies of KZM scaling of work statistics
from an isolated quantum critical point to a critical surface. As
an example, we study the scaling behavior of work statistics in the
2D Kitaev honeycomb model featured with a critical line. By utilizing
the trace formula for quadratic fermionic Hamiltonian, we obtain the
exact characteristic function of work of the 2D Kitaev model at zero
temperature. The results confirm our prediction.
\end{abstract}
\maketitle

\section{Introduction\label{sec:Introduction}}

The dynamics of a nonequilibrium quantum system has received much
attention in recent years, thanks to the development of cold atomic
physics and quantum stimulations$\text{\;}$\citep{Bloch2008RMP,Dziarmaga2010AP,Cazalilla2011rmp,Polkovnikov2011RMP,Altman2015arcmp,Langen2015arcm,Aoki2017rmp,Heyl2018RPP,Monroe2021rmp}.
Generally, a closed system can be brought into nonequilibrium by simply
starting from an initial state which is not an eigenstate of the Hamiltonian,
or changing the parameters of the Hamiltonian. The first case is closely
related to the problem of quantum thermalization, while the second
case, also called quantum quench, will be our main focus in this article.

When the system is driven away from the equilibrium by a time-dependent
Hamiltonian, the usual framework for studying equilibrium system,
namely, the equilibrium statistic mechanics, is inappropriate. Many
physical quantities are proposed to characterize the nonequilibrium
dynamics, such as the density of quasiparticle excitations, correlation
functions, entanglement entropy, fidelity, and nonequilibrium work$\;$\citep{Dziarmaga2010AP,Polkovnikov2011RMP}.
One interesting problem is: is there any universality in the dynamics
of a system following a quench process. And the problem becomes even
more interesting when we consider a system featured with a quantum
phase transition, which does have a universality for the quench across
the quantum critical point (QCP). Typically, a time-dependent many-body
Hamiltonian is extremely hard to penetrate. But in some limits, such
as the fast limit (sudden quench) and the slow limit (adiabatic),
general results are found. These results are based on the fact that
if the response near a QCP dominates the whole dynamics, then we expect
the response should be universal, since the properties near a QCP
are universal. The canonical description of the dynamics from one
gapped phase to another across a QCP, is described by the Kibble-Zurek
mechanism$\;$(KZM)$\;$\citep{Kibble1976jpa,Kibble1980pr,Zurek1985nature,Zurek1996pr}.
The essential of KZM is the adiabatic-impulse-adiabatic approximation.
When the system is far away from the QCP, the gap $\Delta$ is large,
and the response time $\tau\sim\Delta^{-1}$ is small. The system
can easily follow the change of the Hamiltonian. The adiabatic theorem
holds, and the system always stays in the instantaneous ground state
if it is initially prepared in the ground state. However, when the
system is in the vicinity of the QCP, the gap vanishes as $\Delta\sim|\lambda|^{z\nu}$,
and the response time and the correlation length diverge as $\tau\sim|\lambda|^{-z\nu}$,
$\xi\sim|\lambda|^{-\nu}$, where $\lambda$ is a dimensionless parameter
characterizing the deviation from the QCP, $\nu$ is a critical exponent
and $z$ is the dynamic critical exponent$\;$\citep{sachdev2011book}.
Then no matter how slow the quench is, there is a moment at which
the transition rate $|\dot{\lambda}/\lambda|$ is comparable to the
gap $\Delta$. The system fails to keep pace with the change of the
Hamiltonian. The excitation of topological defects is inevitable,
regardless of the quench speed. KZM approximates this diabatic stage
by assuming that the state becomes frozen, until the system is away
from the QCP. After leaving the frozen stage, the system will continue
to evolve quantum adiabatically. The boundary of the frozen stage,
is determined by equaling the transition rate and the gap, i.e., $\Delta(\lambda)\sim|\dot{\lambda}/\lambda|$.
For a linear quench, $\lambda=vt$, $t\in(-\infty,\infty),$ the frozen-out
parameter $\lambda^{*}$ scales as 
\[
\lambda^{*}\sim v^{\frac{1}{\nu z+1}},
\]
and the corresponding frozen-out time $t^{*}\sim v^{-\nu z/(\nu z+1)}$
and a characteristic length scale $\xi^{*}\sim v^{-\nu/(\nu z+1)}$.

Based on the Kibble-Zurek analysis, people have found the scaling
behavior associated with an isolated QCP for the mean value of density
of topological defects$\;$\citep{Dziarmage2005prl,Polkovnikov2005prb,Zurek2005prl},
residual heat$\;$\citep{Polkovnikov2008prl} and entanglement entropy$\;$\citep{Sengupta2009pra},
et al. For example, the density of topological defects is $n_{\mathrm{ex}}\sim(\xi^{*})^{-d}\sim v^{-d\nu/(\nu z+1)}$,
where $d$ is the dimension of the system. Extensions of the KZM include
nonlinear quench$\;$\citep{Sen2008prl,Barankov2008prl}, an anisotropic
QCP$\;$\citep{Amit2010epl,Hikichi2010prb}, multicritical QCPs$\;$\citep{Victor2010epl},
a critical surface$\;$\citep{Sen2008prl,Sengupta2008prl}, and a
gapless phase$\;$\citep{Polkovnikov2008NP}. Recently, higher-order
cumulants beyond the mean value, have attracted much attention. del
Campo used an exactly solvable 1D transverse Ising chain to show that
all cumulants of topological defects exhibit the same scaling behavior$\;$\citep{Campo2018prl}.
Later, Gómez-Ruiz et al. generalized this model-dependent universality
to general topological defects production process, and showed that
the full counting statistics of the topological defects production
is actually universal$\;$\citep{Fernando2020prl}. In parallel, Fei
et al. also showed that aside from the statistics of topological defects,
the work statistics exhibits universality, too$\;$\citep{Fei2020prl}.
Nevertheless, these studies only consider the canonical case when
the system is quenched through an isolated QCP. Here, we extend these
studies by considering a linear quench through a critical surface,
and find distinct scaling behavior of the work statistics, as well
as of the topological defects. We will use the 2D Kitaev honeycomb
model as an example to demonstrate our results.

This article is organized as follows. In Sec. II, we use the adiabatic
perturbation theory to derive the scaling behavior of the work statistics.
In Sec. III, we use the 2D Kitaev honeycomb model to demonstrate the
validity of our main results. In Sec. IV, we discuss our results and
make a summary.

\section{KZM and work statistics Across a quantum Critical Surface}

In this section, we first briefly review the adiabatic perturbation
theory to the KZM$\;$\citep{Polkovnikov2005prb,deGrandi2010book},
and apply it to work statistics. We consider a closed quantum system
under a linear quench, i.e., $\hat{H}(t)=\hat{H}(\lambda(t))$, where
$\lambda(t)=vt$, $t\in(t_{0},t_{1})$ is the work parameter. The
initial and final values are denoted as $\lambda_{0}\equiv vt_{0}$,
and $\lambda_{1}\equiv vt_{1}$, respectively. We assume the low energy
excitation near the QCP can be described by quasiparticles, and the
dispersion relation is $\omega_{k}=c|k|^{z}$$\;$\citep{Halperin2019pt}.
Here $c$ is a constant and $k$ denotes the momentum. And we further
adapt the single-particle excitation approximation$\;$\citep{Fei2020prl}
that at most one quasiparticle can be excited in every mode$\;$\footnote{For the fermionic quasiparticle, it is always satisfied, while for
bosonic excitation, this approximation is fulfilled only for small
excitation probability.}.

For a time dependent Hamiltonian $\hat{H}(t)$, we first expand the
state $\ket{\psi(t)}$ in terms of the instantaneous eigenstate $\ket{u_{n}(t)}$
\[
\ket{\psi(t)}=\sum_{n}a_{n}(t)|u_{n}(t)\rangle.
\]
From the Schr\"{o}dinger equation $i\partial_{t}|\psi\rangle=\hat{H}(t)|\psi\rangle,$
the $n$- th coefficient $a_{n}(t)$ satisfies the following equation
\begin{equation}
i\partial_{t}a_{n}(t)+i\sum_{m}a_{m}(t)\langle u_{n}|\partial_{t}|u_{m}\rangle=E_{n}(t)a_{n}(t),\label{eq:coeff_a-1}
\end{equation}
where $E_{n}(t)$ is the instantaneous eigenenergy of the Hamiltonian
$\hat{H}(t)$ corresponding to the eigenstate $\ket{u_{n}(t)}.$ And
we do a gauge transformation 
\begin{equation}
\tilde{a}_{n}(t)=e^{-i\gamma_{n}(t)}a_{n}(t)\label{eq:a_gaug}
\end{equation}
where $\gamma_{n}(t)=\int_{t_{0}}^{t}dt'E_{n}(t')$. Then Eq.$\;$(\ref{eq:coeff_a-1})
becomes
\[
\partial_{t}\tilde{a}_{n}(t)=-\sum_{m}\tilde{a}_{m}(t)\langle u_{n}|\partial_{t}|u_{m}\rangle e^{i(\gamma_{n}(t)-\gamma_{m}(t))}.
\]
 The solution is 
\begin{align}
\tilde{a}_{n}(t) & =-\sum_{m}\int_{t_{0}}^{t}dt'\tilde{a}_{m}(t')\langle u_{n}|\partial_{t'}|u_{m}\rangle e^{i(\gamma_{n}(t')-\gamma_{m}(t'))}\nonumber \\
 & =-\sum_{m}\int_{\lambda_{0}}^{\lambda}d\lambda'\tilde{a}_{m}(\lambda')\langle u_{n}|\partial_{\lambda'}|u_{m}\rangle e^{i(\tilde{\gamma}_{n}(\lambda')-\tilde{\gamma}_{m}(\lambda'))},\label{eq:a_tilde_eq}
\end{align}
where $\tilde{\gamma}_{n}(\lambda)=v^{-1}\int_{-\infty}^{\lambda}d\lambda'\;E_{n}(\lambda')$.
In the quasiparticle picture, the eigenstate $\ket{u_{n}}$ can be
decomposed into independent $k$ modes. Under the single-particle
approximation, if the initial state is the ground state of $\hatH(\lambda(t_{0}))$,
the probability $p_{k}=|\tilde{a}_{k}|^{2}$ for exciting one quasiparticle
in mode $k$ with energy $\omega_{k}$ at the end of the protocol
can be written as$\;$\citep{Dziarmaga2010AP}
\begin{equation}
p_{k}\approx\int_{\lambda_{0}}^{\lambda_{1}}d\lambda'\left|\langle1_{k}|\partial_{\lambda'}|0_{k}\rangle e^{iv^{-1}\int_{-\infty}^{\lambda'}d\lambda''\omega_{k}(\lambda'')}\right|^{2}.\label{eq:prob}
\end{equation}
Near the QCP, the general scaling argument implies that 
\begin{equation}
\omega_{k}(\lambda)=\lambda{}^{z\nu}F(\frac{k}{\lambda^{\nu}}),\quad\bra{1_{k}}\partial_{\lambda}\ket{0_{k}}=\frac{1}{\lambda}G(\frac{k}{\lambda^{\nu}})\label{eq:scale-arg}
\end{equation}
 where $F(x)$ and $G(x)$ are two non-universal functions with universal
asymptotic expression $F(x)\propto x^{z}$ and $G(x)\propto1/x^{1/\nu}$
for $x\gg1$, due to the fact that high momentum spectrum does not
dependent on $\lambda$$\;$\citep{Dziarmaga2010AP,deGrandi2010book}.
We change the variable 
\[
\eta=\lambda k^{-1/\nu}
\]
 and substitute Eq.$\;$(\ref{eq:scale-arg}) into Eq.$\;$(\ref{eq:prob}).
We obtain
\begin{equation}
p_{k}\approx\left|\int d\eta\frac{1}{\eta}G(\frac{1}{\eta^{\nu}})e^{i\frac{k^{1/\nu+z}}{v}\int_{-\infty}^{\eta}d\eta'F(1/\eta')\eta'^{z\nu}}\right|^{2},\label{eq:prob_2}
\end{equation}
which depends on $k$ only through the dimensionless combination $k^{(1+z\nu)/\nu}/v$.
It implies that there exists a characteristic length scale

\[
\xi^{*}\sim(k^{*})^{-1}\simeq v^{-\frac{\nu}{z\nu+1}}
\]
 which is consistent with the prediction of KZM.

Now we consider the work statistics. We start with the initial density
matrix $\hat{\rho}_{0}$. The dynamics is governed by the Hamiltonian
$\hatH(t)$. We adapt the two-point measurement scheme, i.e., we measure
the instantaneous eigenenergy at initial time $t=t_{0}$ and final
time $t=t_{1}$. The probability distribution of work is 
\[
p(w)=\sum_{n,m}\delta\left[w-(E_{n}'-E_{m})\right]p_{nm}\bra{u_{m}}\hat{\rho}_{0}\ket{u_{m}},
\]
 where $\ket{u_{m}}(|u_{n}'\rangle)$ is the measured eigenstate of
$\hat{H}(t_{0})(\hat{H}(t_{1}))$ corresponding to the eigenenergy
$E_{m}(E'_{n})$ . The transition probability is $p_{nm}=|\bra{u'_{n}}\hat{U}(t_{1},t_{0})\ket{u_{m}}|^{2}$
with the evolution operator $\hat{U}(t_{1},t_{0})=\mcT e^{-i\int_{t_{0}}^{t_{1}}\hat{H}(t)dt}$
where $\mcT$ is the time-ordering operator. The energy difference
$E'_{n}-E_{m}$ is defined as the fluctuating work for one realization.
It is more convenient to consider the characteristic function of work$\;$(CFW),
defined as the Fourier transform of the probability distribution $p(w)$
\begin{align}
Z(\chi) & =\int_{-\infty}^{\infty}dwe^{i\chi w}p(w)\nonumber \\
 & =\tr[\cohU(t_{1},t_{0})e^{i\chi\hatH(t_{1})}\hat{U}(t_{1},t_{0})e^{-i\chi\hat{H}(t_{0})}\hat{\rho}_{0}].\label{eq:cfw}
\end{align}
The cumulant generating function$\;$(CGF) is defined as the logarithm
of the CFW and allows a series expansion around $\chi=0$
\[
\mcG(\chi)\equiv\frac{1}{N}\ln Z(\chi)=\sum_{n=1}^{\infty}\frac{(i\chi)^{n}}{n!}\kappa_{n},
\]
 where $\kappa_{n}\equiv(-i)^{n}\partial^{n}\mcG(\chi)/\partial\chi^{n}$
is the $n$-th cumulant of work.

In the quasiparticle picture, the statistics of excitations can be
described by a combination of $N$ Bernoulli trials associated with
the probability $p_{k}$ of exciting a quasiparticle in mode $k$$\;$\citep{Fei2020prl,Fernando2020prl,Campo2018prl}.
And the work statistics is related to the statistics of excitation
by the fact that the excitation of a quasiparticle in mode $k$ requires
an amount of excess work $w(k)=\omega_{k}$, since the initial state
is the ground state which hosts no quasiparticle. Here we drop out
the work associated with the change of zero-point energy. The CFW
and the CGF take the following form approximately
\begin{align}
Z(\chi) & \sim\prod_{k}(1-p_{k}+p_{k}e^{i\chi w(k)}),\\
\mcG(\chi) & \sim\int\frac{d^{d}k}{(2\pi)^{d}}\ln(1-p_{k}+p_{k}e^{i\chi w(k)}).
\end{align}
The first two cumulants of work $\kappa_{1}$ and $\kappa_{2}$ are
the mean value and the variance 
\begin{align*}
\kappa_{1} & =\avg w\sim\int d^{d}kp_{k}w(k)=\int d^{d}k\kappa_{1}(k),\\
\kappa_{2} & =\avg{w^{2}}-\avg w^{2}\sim\int d^{d}kp_{k}(1-p_{k})w^{2}(k)=\int d^{d}k\kappa_{2}(k).
\end{align*}
Higher-order cumulant $\kappa_{n}$$\;$($n>2$) can be obtained from
the recursion relation of binomial distribution for every $k$ mode
\[
\kappa_{n+1}(k)=w(k)p_{k}(1-p_{k})\frac{d\kappa_{n}(k)}{dp_{k}}.
\]
For example, the third cumulant$\;$(also called skewness) is 
\[
\kappa_{3}\sim\int d^{d}kp_{k}(1-p_{k})(1-2p_{k})w^{3}(k).
\]
The nonzero higher-order cumulant signals a non-Gaussian distribution.
We see that the work statistics is quiet similar to the full counting
statistics of topological defects$\;$\citep{Fernando2020prl}. In
the two limits $\lambda(t_{0})=-\infty,\lambda(t_{1})=\infty$ separated
by an isolated QCP, the energy $\omega_{k}$ becomes $k$-independent
asymptotically. So for a linear quench, the scaling behavior for the
work statistics is found to be the same as that of the topological
defects, namely, 
\begin{equation}
\kappa_{n}\sim\begin{cases}
v^{d\nu/(z\nu+1)} & d\nu/(z\nu+1)<2\\
v^{2}\ln v & d\nu/(z\nu+1)=2\\
v^{2} & d\nu/(z\nu+1)>2
\end{cases},\label{eq:fei_work_scal}
\end{equation}
 which was obtained in Ref.$\;$\citep{Fei2020prl}.

In the above discussion, we focus on the situation where the system
has an isolated QCP only. In the following, we will consider a more
general situation: the system is featured with a $(d-m)$-dimensional
critical surface. In this situation, it has been found that the mean
density of the topological defects scales as $n_{\mathrm{ex}}\sim v^{m\nu/(z\nu+1)}$
, which generalizes the original KZM scaling behavior$\;$\citep{Sengupta2008prl,Mondal2008prb}.
According to Refs.$\;$\citep{Sengupta2008prl,Mondal2008prb}, the
existence of a $(d-m)$-dimensional critical surface reduces the available
phase space from $\Omega\sim k^{d}$ to $\Omega\sim k^{m}.$ From
the adiabatic perturbation theory, the leading term of the work cumulant
$\kappa_{n}$ is 
\begin{align}
\kappa_{n} & \sim\int d^{m}kp_{k}w^{n}(k)\nonumber \\
 & =\int d^{m}k\;w^{n}(k)\int_{-\infty}^{\infty}d\lambda'\langle1_{k}|\partial_{\lambda'}|0_{k}\rangle e^{iv^{-1}\int_{-\infty}^{\lambda'}d\lambda''\omega_{k}(\lambda'')}|^{2}.\label{eq:n_work_cum_per}
\end{align}
We introduce a new pair of variables $(\zeta,\phi)$
\[
\lambda=\zeta v^{1/(z\nu+1)},\quad k=\phi v^{\nu/(z\nu+1)}.
\]
 Then Eq.$\;$(\ref{eq:n_work_cum_per}) becomes 
\begin{equation}
\kappa_{n}\sim v^{m\nu/(z\nu+1)}\int d^{m}\phi\;\tilde{w}^{n}(\phi)K(\phi),\label{eq:kapp_n}
\end{equation}
where $\tilde{w}(\phi)=w(\phi v^{\nu/(z\nu+1)})$ and 
\[
K(\phi)=\left|\int_{-\infty}^{\infty}\frac{d\zeta}{\zeta}\;G(\frac{\phi}{|\zeta|^{\nu}})e^{i\int_{-\infty}^{\zeta}d\zeta'|\zeta'|^{z\nu}F(\frac{\phi}{|\zeta|^{\nu}})}\right|^{2}.
\]
The scaling behavior of $\kappa_{n}$ is determined by the integral
in Eq.$\;$(\ref{eq:kapp_n}). In the following, we will examine the
convergence of the integral. We consider the integration domain $\bar{S}$
for a large $\phi$ where the asymptotic expressions of $F(x)$ and
$G(x)$ can be applied. The integral in $\bar{S}$ is 
\begin{align}
 & \int_{\bar{S}}d^{m}\phi\;\tilde{w}^{n}(\phi)K(\phi)\nonumber \\
 & =\int_{\bar{S}}d^{m}\phi\;\tilde{w}^{n}(\phi)\left|\int_{-\infty}^{\infty}d\zeta\;\frac{\mathrm{sign(\zeta)}}{\phi^{1/\nu}}e^{i\int_{-\infty}^{\zeta}d\zeta'\phi^{z}}\right|^{2}\\
 & \sim\int_{\bar{S}}d^{m}\phi\;\tilde{w}^{n}(\phi)\frac{1}{\phi^{2(1+\nu z)/\nu}}.\label{eq:high_phi_int}
\end{align}
If $m\nu/(1+\nu z)>2$, the integral in Eq.$\;$(\ref{eq:high_phi_int})
diverges, and the contributions from the high energy modes dominate,
which further implies the breakdown of adiabatic perturbation theory.
In this case, the scaling is quadratic, given by the regular analytic
adiabatic perturbation theory$\;$\citep{deGrandi2010book}. For the
critical case $m\nu/(1+\nu z)\text{=2},$ a logarithm correction is
expected. If $m\nu/(1+\nu z)<2$, the integral in Eq.$\;$(\ref{eq:high_phi_int})
converges, and the scaling behavior of $\kappa_{n}$$\;$(Eq.$\;$(\ref{eq:kapp_n}))
is given by $v^{m\nu/(z\nu+1)}$. So we conclude that the $n$-th
work cumulant also exhibits a different scaling behavior from Eq.$\;$(\ref{eq:fei_work_scal}),
i.e,
\begin{equation}
\kappa_{n}\sim\begin{cases}
v^{m\nu/(z\nu+1)} & m\nu/(z\nu+1)<2\\
v^{2}\ln v & m\nu/(z\nu+1)=2\\
v^{2} & m\nu/(z\nu+1)>2
\end{cases}.\label{eq:scal_work_crisur}
\end{equation}
 Eq.$\;$(\ref{eq:scal_work_crisur}) is the main result of our article.
In the following we will demonstrate the validity of this result with
an exactly solvable model, the 2D Kitaev honeycomb model.

\section{Example: the 2D Kiteav Honeycomb Model}

Our result Eq.$\;$(\ref{eq:scal_work_crisur}) is quite general.
The usual quantum phase transition systems featured with an isolated
QCP are included as a special case. For example, the scaling behavior
of 1D transverse Ising model considered in Refs.$\;$\citep{Campo2018prl,Fei2020prl}
is already included in Eq.$\;$(\ref{eq:scal_work_crisur}), since
an isolated QCP can be seen as a critical surface of dimension $d-m=0$,
so $m=d$. From Eq.$\;$(\ref{eq:scal_work_crisur}), we recover Eq.$\;$(\ref{eq:fei_work_scal}).
Besides these systems, another example featured with a 1D critical
surface is the 2D Kitaev honeycomb model$\;$(KHM)$\;$\citep{Kitaev2006np,Mondal2008prb,Sengupta2008prl},
of which the ground state is exactly solvable. It is also worth mentioning
that the KHM, which is initially proposed as a toy model to demonstrate
the physics of quantum spin liquid, has been possibly realized in
experiments$\;$\citep{Trebst2017arxiv,Zhou2017rmp,Takagi2019nrp,Broholm2020science}.
The promising material candidates realizing KHM include $\alpha$-$\mathrm{RuCl_{3}}$$\;$\citep{Sandilands2015prl,Banerjee2016nm,Do2017nature,Zheng2017prl,Baek2017prl}
and $\mathrm{(Na_{1-x}Li_{x})_{2}IrO_{3}}$$\;$\citep{Manni2014prb,Hwan2015np,Das2019prb}.

Previous studies in quench dynamics of KHM usually focus on the mean
value of physical quantities, such as the density of topological defects,
residue heat, and correlation functions$\;$\citep{Sengupta2008prl,Mondal2008prb,Hikichi2010prb,Patel2012orb}.
Here we use the trace formula developed in Ref.$\;$\citep{Fei2019prr}
for quadratic fermionic models to calculate the CFW associated with
linearly quenching KHM, thus generalizing the previous studies from
the mean value to all cumulants of work.

\begin{figure}[h]
\includegraphics[scale=0.5]{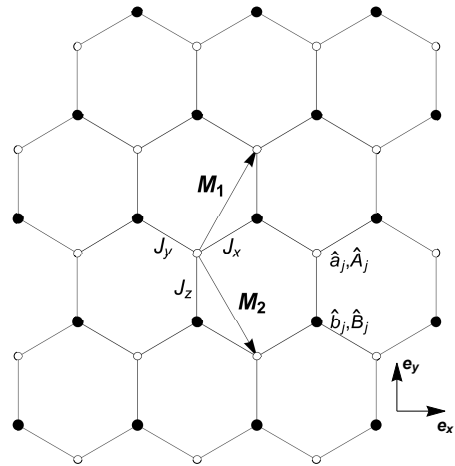} \caption{The honeycomb lattice. Empty~(filled) circles form the sub-lattice
$a$~$(b)$. The bonds with positive~(negative) slope are called
$x$~$(y)$ bonds. The vertical bonds are called $z$ bonds. $\mathbf{e_{x}}$
and $\mathbf{e_{y}}$ denote the unit vectors along the horizontal
and the vertical directions.}
\label{lattice}
\end{figure}

\begin{figure}
\includegraphics[scale=0.55]{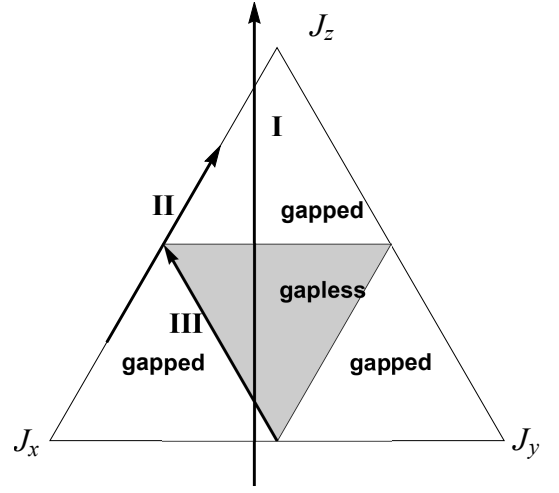}

\caption{The phase diagram of Kitaev model in the plane $J_{x}+J_{y}+J_{z}=1$,
$J_{x},J_{y},J_{z}\protect\geq0$. The shadowed triangle indicates
the gapless phase of the KHM. The three arrows stand for three quench
protocols: {\color{blue}(I)$\;$$J_{x}=\cos(\pi/3)$, $J_{y}=\sin(\pi/3)$; (II)$\;$$\;J_{x}=1,J_{y}=0$;
(III)$\;$$\;J_{x}=J_{y}+J_{z}=1.$} In all three protocols, we linearly
vary $J_{z}=vt$ which starts from $J_{0}<-1$ to $J_{1}>1$.\label{fig:The-phase-diagram}}
\end{figure}

The Hamiltonian of the KHM is 
\begin{align}
H & =\sum_{x,\avg{jl}}J_{x}\sigma_{ja}^{x}\sigma_{lb}^{x}+\sum_{y,\avg{jl}}J_{y}\sigma_{ja}^{y}\sigma_{lb}^{y}+\sum_{z,\avg{jl}}J_{z}\sigma_{ja}^{z}\sigma_{jb}^{z},\label{KMH}
\end{align}
where $\sigma^{x,y,z}$ is the Pauli matrix, $J_{x},J_{y},J_{z}$
denote the coupling strength, and $\avg{jl}$ denotes the nearest
neighbors. The lattice is shown in Fig.~\ref{lattice}. 
We adapt the Jordan-Wigner transformation to introduce Majorana fermions
$\aoa,\aoA$~($\aob,\aoB$) for sub-lattice a~(b), namely, 
\begin{align}
\aoa_{j} & =\lrb{\prod_{k<j}\sigma_{k}^{z}}\sigma_{j}^{y},\quad\hat{A}_{j}=\lrb{\prod_{k<j}\sigma_{k}^{z}}\sigma_{j}^{x},\label{eq:a_A_JW}\\
\aob_{j} & =\lrb{\prod_{k<j}\sigma_{k}^{z}}\sigma_{j}^{x},\quad\hat{B}_{j}=\lrb{\prod_{k<j}\sigma_{k}^{z}}\sigma_{j}^{y}.\label{eq:b_B_JW}
\end{align}
The product in Eqs.$\;$(\ref{eq:a_A_JW}, \ref{eq:b_B_JW}) is over
all sites on the one-dimensional contour which threads the entire
lattice$\;$\citep{Feng2007prl,Chen2008jpa}. It can be checked that
they indeed anticommute with each other and satisfy the Majorana condition,
for example, 
\[
\lrc{\aoa_{j},\aoA_{l}}=\lrc{\aoa_{j},\hat{b}_{l}}=\lrc{\aoa_{j},\hat{B}_{l}}=0,\lrc{\aoa_{j},\aoa_{l}}=2\delta_{jl}.
\]
We rewrite the Hamiltonian in the form of Majorana fermions 
\[
\hat{H}=\sum_{j}i\lrb{J_{x}\aob_{j}\aoa_{j-\bfM_{1}}+J_{y}\aob_{j}\aoa_{j+\bfM_{2}}+J_{z}\hat{D}_{j}\hat{b}_{j}\aoa_{j}},
\]
where the lattice vectors are $\bfM_{1}=\frac{\sqrt{3}}{2}{\bf e_{x}}+\frac{3}{2}{\bf e_{y}}$,
$\bfM_{2}=\frac{\sqrt{3}}{2}{\bf e_{x}}-\frac{3}{2}{\bf e_{y}}$,
and $\hatD_{j}\equiv-i\aoA_{j}\aoB_{j}$. It can be shown that $\hatD_{j}$
is a constant of motion, i.e., commutes with $\hatH$$\;$\citep{Feng2007prl,Chen2008jpa}.
Since $\hatD_{j}^{2}=1$, $\hatD_{j}=\pm1$ in its eigenspaces. According
to the Lieb's theorem$\;$\citep{Lieb1994prl}, the ground state is
in the sector with all $\hatD_{j}=1$; a negative $\hatD_{j}$ corresponds
to a topological excitation$\;$\citep{Kitaev2006np,Feng2007prl,Chen2008jpa}.
Since we will focus on the ground state, we simply set all $\hatD_{j}$
to be 1. The Hamiltonian in the so-called zero-flux sector is 
\begin{align}
\hat{H}=\sum_{j}i\lrb{J_{x}\aob_{j}\aoa_{j-{\bf M_{1}}}+J_{y}\aob_{j}\aoa_{j+{\bf M_{2}}}+J_{z}\aob_{j}\aoa_{j}}.
\end{align}
Now we apply Fourier transform 
\begin{align*}
\aoa_{j} & =\sqrt{\frac{2}{N_{1}N_{2}}}\sum_{k\in\mathrm{HBZ}}\lrp{e^{ikj}\aoa_{k}+e^{-ikj}\coa_{k}},\\
\aob_{j} & =\sqrt{\frac{2}{N_{1}N_{2}}}\sum_{k\in\mathrm{HBZ}}\lrp{e^{ikj}\aob_{k}+e^{-ikj}\cob_{k}},
\end{align*}
with the fermionic operator $\aoa_{k},$ $\aob_{k}$ for the $k$-th
mode. The Brillouin zone$\;$(BZ) is spanned by $\bfk=2\pi\frac{p_{1}}{N_{1}}\bfk_{1}+2\pi\frac{p_{2}}{N_{2}}\bfk_{2},$
where $p_{1}\in\lrp{-\frac{N_{1}}{2},\frac{N_{1}}{2}},p_{2}\in\lrp{-\frac{N_{2}}{2},\frac{N_{2}}{2}}$
and $\bfk_{1}=\frac{1}{\sqrt{3}}{\bf e_{x}}+\frac{1}{3}{\bf e_{y}},\bfk_{2}=\frac{1}{\sqrt{3}}{\bf e_{x}}+\frac{-1}{3}{\bf e_{y}}$.
In the above definition, we have assumed that there are $N_{1}$ empty$\;$(filled)
sites in every column and $N_{2}$ in every row. And we restrict the
momentum in the left half Brillouin zone$\;$(HBZ) satisfying $\bfk\cdot{\bf e_{x}\geq0}$.
The Hamiltonian is transformed into 
\begin{align}
\hat{H} & =2\sum_{k\in\mathrm{HBZ}}\begin{pmatrix}\coa_{k} & \cob_{k}\end{pmatrix}\begin{pmatrix}0 & -if\\
if^{*} & 0
\end{pmatrix}_{k}\begin{pmatrix}\aoa_{k}\\
\aob_{k}
\end{pmatrix}\\
 & =\half\sum_{k\in\mathrm{HBZ}}\hat{\alpha}_{k}^{T}\mcH_{k}\hat{\alpha}_{k},
\end{align}
where 
\[
\hat{\alpha}_{k}^{T}=\begin{pmatrix}\aoa & \aob & \coa & \cob\end{pmatrix}_{k},\quad f_{k}=J_{x}e^{i\bfk\cdot\bfM_{1}}+J_{y}e^{-i\bfk\cdot\bfM_{2}}+J_{z}
\]
with 
\begin{equation}
\mcH_{k}=2\begin{pmatrix}0 & 0 & 0 & -if^{*}\\
0 & 0 & if & 0\\
0 & -if & 0 & 0\\
if^{*} & 0 & 0 & 0
\end{pmatrix}_{k}.\label{eq:ham_k_HBZ}
\end{equation}
The energy spectrum is 
\begin{align}
E_{k} & =\pm2\sqrt{\epsilon_{k}^{2}+\Delta_{k}^{2},}
\end{align}
where 
\begin{align}
\epsilon_{k} & =\mathrm{Re}\lrb{f_{k}}=J_{x}\cos(\bfk\cdot\bfM_{1})+J_{y}\cos(\bfk\cdot\bfM_{2})+J_{z},\\
\Delta_{k} & =\mathrm{Im}\lrb{f_{k}}=J_{x}\sin(\bfk\cdot\bfM_{1})-J_{y}\sin(\bfk\cdot\bfM_{2}).
\end{align}
The model is gapless when $J_{x},J_{y}$ and $J_{z}$ satisfy the
triangular inequality$\;$(see Fig.$\;$\ref{fig:The-phase-diagram})
\[
|J_{x}|\leq|J_{y}|+|J_{z}|;\quad|J_{y}|\leq|J_{x}|+|J_{z}|;\quad|J_{z}|\leq|J_{x}|+|J_{z}|.
\]

We consider the following linear quench protocol, 
\begin{equation}
J_{x}=\cos\theta,\quad J_{y}=\sin\theta,\quad J_{z}=vt,\label{eq:procotol}
\end{equation}
where $\theta$ is fixed, and $J_{z}$ is quenched linearly from $J_{0}=vt_{0}$
to $J_{1}=vt_{1}$ with $J_{0}\ll-1$ and $J_{1}\gg1$. The phase
diagram in the plane $J_{x}+J_{y}+J_{z}=1$ and the quench protocol
are shown in Fig.$\;$\ref{fig:The-phase-diagram}.

The CFW$\;$(\ref{eq:cfw}) can be rewritten as 
\begin{align*}
Z(\chi) & =\tr[\cohU(t_{1},t_{0})e^{i\chi\hatH(t_{1})}\hat{U}(t_{1},t_{0})e^{-i\chi\hat{H}(t_{1})}\hat{\rho}_{0}]\\
 & =\tr[e^{i\chi\hatH^{H}(t_{1})}e^{-(i\chi+\beta)\hat{H}(t_{0})}],
\end{align*}
where 
\begin{align}
\hatH^{H}(t_{1}) & =\cohU(t_{1},t_{0})\hatH(t_{1})\hat{U}(t_{1},t_{0})\nonumber \\
 & =\sum_{k\in\mathrm{HBZ}}\hat{\alpha}_{k}^{T}(t_{1})\mcH_{k}(t_{1})\hat{\alpha}_{k}(t_{1})\label{eq:Ham_final}
\end{align}
 is the final Hamiltonian in the Heisenberg picture, and the ground
state is represented by a thermal equilibrium state $\hat{\rho}_{0}=e^{-\beta\hat{H}(t_{0})}/\tr[e^{-\beta\hat{H}(t_{0})}]$
with $\beta\to\infty$. $\hat{\alpha}_{k}(t_{1})$ is determined by
the Heisenberg equation of motion
\[
i\frac{d}{dt}\hat{\alpha}_{k}^{H}(t)=\lrb{\hat{\alpha}_{k}^{H},\hat{H}^{H}(t)}.
\]
The solution is linear due to the quadratic form of the Hamiltonian
\[
\hat{\alpha}_{k}^{H}(t)=L_{k}(t)\hat{\alpha}_{k}+C_{k}(t).
\]
where the coefficient matrices $L_{k}(t)$ and $C_{k}(t)$ satisfy
\begin{align}
i\frac{d}{dt}L_{k}(t) & =\tau_{F}\mcH_{k}(t)L_{k}(t),\quad L(t_{0})=\bfI_{4\times4},\nonumber \\
i\frac{d}{dt}C_{k}(t) & =\tau_{F}\mcH_{k}(t)C_{k}(t),\quad C(t_{0})=0,\label{eq:L and C}
\end{align}
 with 
\[
\tau_{F}=\begin{pmatrix}0 & \bfI_{2\times2}\\
\bfI_{2\times2} & 0
\end{pmatrix}
\]
 and $\bfI_{n\times n}$ is the $n$-dimensional identity matrix.
The dynamics is mapped to the standard Landau-Zener problem$\;$\citep{deGrandi2010book}.
We find $C_{k}(t_{1})=0$, and the transition matrix $L_{k}(t_{1})=\mathrm{diag}(\Lambda_{k},\Lambda_{k}^{*})$
is a block diagonal matrix where the $2\times2$ matrix $\Lambda_{k}$
in the limit $t_{0}\to-\infty$, $t_{1}\to\infty$ is 
\begin{align}
\small\Lambda_{k}(t_{1}) & \approx\sqrt{p_{k}}\left[\cos\Theta_{k}(t_{1},t_{0})\bfI_{2\times2}+i\sin\Theta_{k}(t_{1},t_{0})\sigma^{y}\right]\nonumber \\
 & \quad-i\sqrt{1-p_{k}}\left[\cos\Theta'_{k}(t_{1},t_{0})\sigma^{z}-\sin\Theta'_{k}(t_{1},t_{0})\sigma^{x}\right],\label{eq:lam_upp}
\end{align}
where $\Theta(t_{1},t_{0})$ and $\Theta'(t_{1},t_{0})$ are two phase
factors which do not appear in the final result of the CFW in the
above limits. For completeness, we give the expression of $\Theta(t_{1},t_{0})$
and $\Theta'(t_{1},t_{0})$ in Appendix$\;$\ref{sec:Landau-Zener-Problem-of}.
The Landau-Zener transition probability $p_{k}$ for mode $k$ is
\begin{equation}
p_{k}=e^{-\frac{2\pi}{v}\Delta_{k}^{2}}=e^{-\frac{2\pi}{v}\left[J_{x}\sin(\bfk\cdot\bfM_{1})-J_{y}\sin(\bfk\cdot\bfM_{2})\right]^{2}}.\label{eq:prob_KHM}
\end{equation}
Substituting these results into Eq.$\;$(\ref{eq:Ham_final}), the
Hamiltonian at the final time can be written as
\begin{align*}
\hat{H}^{H}(t_{1}) & =\sum_{k\in\mathrm{HBZ}}\hat{\alpha}_{k}^{T}(t_{1})\mcH_{k}(t_{1})\hat{\alpha}_{k}(t_{1})\\
 & =\sum_{k\in\mathrm{HBZ}}\hat{\alpha}_{k}^{T}L_{k}^{T}(t_{1})\mcH_{k}(t_{1})L_{k}(t_{1})\hat{\alpha}_{k}.
\end{align*}
 Now we use the trace formula$\;$\citep{Fei2019prr}
\begin{equation}
\small\tr\lrb{\prod_{k}\exp\lrp{\half\hat{\alpha}_{k}^{T}\mcH_{k}\hat{\alpha}_{k}}}=\lrc{\det\lrb{\prod_{k}\exp\lrp{\tau_{F}\mcH_{k}}+I}}^{\frac{1}{2}}\label{eq:trace-formula}
\end{equation}
and find that the CFW at zero temperature can be approximately$\;$(exactly
in the limit $t_{0}\to-\infty,$ $t_{1}\to\infty$) written as 
\begin{align}
Z(\chi) & =\prod_{k}Z(\chi,k)\nonumber \\
 & =e^{-2i(J_{1}+J_{0})\chi}\prod_{k}\left[(1-p_{k})+p_{k}e^{4iJ_{1}\chi}\right].\label{eq:cfw_calulated}
\end{align}
We see that the CFW consists of two parts: a global work corresponding
to the shift of zero-point energy$\;$(the ground state energies are
$2J_{0}$ and $-2J_{1}$ at the initial time $t_{0}$ and the final
time $t_{1}$, respectively); a product of contributions from different
$k$ modes. For every $k$ mode, there is a probability $p_{k}$ of
exciting a quasiparticle at the cost of work $4J_{1}$. The CGF can
be expressed as 
\begin{align}
\mcG(\chi) & =\frac{1}{N}\ln Z(\chi)=\int_{\mathrm{HBZ}}\frac{d^{2}k}{S_{\mathrm{BZ}}}\ln Z(k,\chi)\nonumber \\
 & \approx\int_{\mathrm{HBZ}}\frac{d^{2}k}{S_{\mathrm{BZ}}}\left[\ln\lrp{1-p_{k}+p_{k}e^{4iJ_{1}\chi}}-2i\chi(J_{0}+J_{1})\right],\label{eq:CGF_KHM}
\end{align}
where $S_{\mathrm{BZ}}$ is the area of the BZ. As mentioned before,
the second term in Eq.$\;$(\ref{eq:CGF_KHM}) only shifts the mean
value of work $\kappa_{1}$ by a constant, and does not affect higher-order
cumulants $\kappa_{n}\;(n\geq2)$. So we can ignore it in $\kappa_{1}$
in the analysis of the scaling behavior. Substituting Eq.$\;$(\ref{eq:prob_KHM})
into Eq.$\;$(\ref{eq:CGF_KHM}), we obtain the scaling behavior of
the first three cumulants
\begin{align}
\kappa_{1} & =\int_{\mathrm{HBZ}}\frac{d^{2}k}{S_{\mathrm{BZ}}}4J_{1}p_{k}\sim v^{1/2},\nonumber \\
\kappa_{2} & =\int_{\mathrm{HBZ}}\frac{d^{2}k}{S_{\mathrm{BZ}}}(4J_{1})^{2}p_{k}(1-p_{k})\sim v^{1/2},\nonumber \\
\kappa_{3} & =\int_{\mathrm{HBZ}}\frac{d^{2}k}{S_{\mathrm{BZ}}}(4J_{1})^{3}p_{k}(1-p_{k})(1-2p_{k})\sim v^{1/2},\label{eq:first_three_cum}
\end{align}
which are consistent with Eq.$\;$(\ref{eq:scal_work_crisur}) with
$m=\nu=z=1$ for the KHM. Higher-order cumulants of work can be obtained
in a similar way.

\begin{figure}
\includegraphics[scale=0.4]{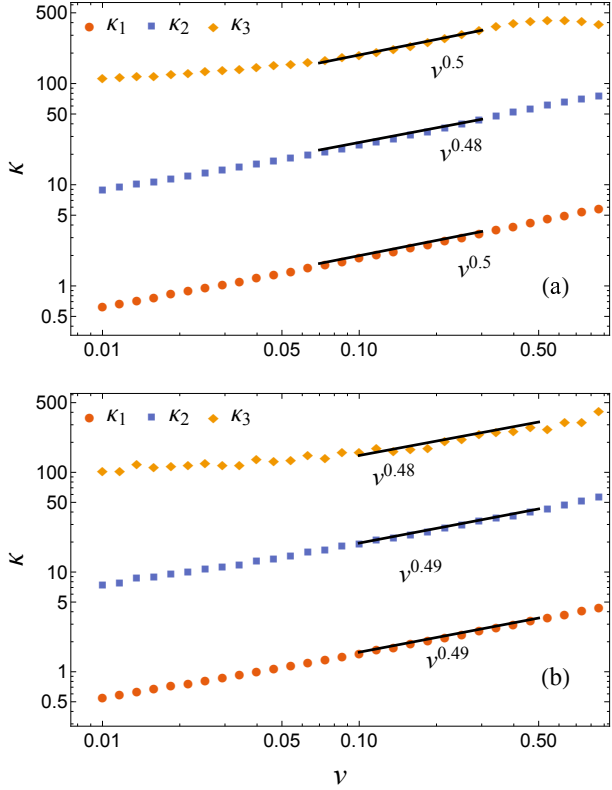}

\caption{The first three cumulants of work for the KHM quenched linearly across
a gapless phase. The parameters are chosen to be $J_{1}=10,J_{0}=-10$,
$\theta=\pi/3$ {\color{blue}for (a) and $\theta=0$ for (b)}. {\color{blue}The markers} correspond
to the results from numerical integration of the time-dependent Schr$\ddot{\mathrm{o}}$dinger
equations in momentum space and solid lines are the fits. The site
number is $400\times400$. The {\color{blue}slopes} of the solid lines are in good
agreement with the theoretical prediction (\ref{eq:scal_work_crisur}).
\label{fig:The-first-three}}
\end{figure}

In the following, we consider two quench protocols$\;$(a third protocol
that quenches along one edge of the gapless phase is left in Appendix$\;$\ref{sec:Quench-through-one-edge}).
Although they all lead to the same scaling behaviors, the details
differ. In protocol I, we choose $\theta=\pi/3$$\;${[}see Fig.$\;$\ref{fig:The-phase-diagram}$\;$(I){]}.
The system crosses through the gapless phase without touching any
vertexes of the shaded triangle in Fig.$\;$\ref{fig:The-phase-diagram}.
Corresponding to any points in the gapless phase, there are some isolated
Fermi points in the HBZ for which the gap vanishes. As $J_{z}$ ramps
up, these isolated Fermi points draw critical lines in the HBZ. In
protocol II, $\theta=0$$\;${[}see Fig.$\;$\ref{fig:The-phase-diagram}$\;$(II){]},
we quench the system along one edge of the big triangle in Fig.$\;$\ref{fig:The-phase-diagram}.
In this protocol, only one (multi)critical point is touched. At this
(multi)critical point, the gap vanishes along a line$\;$(instead
of some isolated points) in the HBZ. Now we numerically integrate
the time-dependent Schr$\ddot{\mathrm{o}}$dinger equations in momentum
space, and show the simulation results of cumulants of work statistics
in Fig.$\;$\ref{fig:The-first-three}. It can be seen that the numerical
results agree very well with our theoretical prediction.

\section{Discussion And Summary}

Before concluding our article, we would like to give the following
remarks: (1)$\;$We consider the scaling behavior of the cumulants
of the work distribution only. In parallel, the scaling behavior of
the cumulants of the topological defects can be studied in a similar
way. In fact, they exhibit exactly the same scaling behaviors. (2)$\;$The
critical surface can be easily found in a wide range of models. It
either has a gapless phase while the gap vanishes in a handful of
isolated points in the BZ, or it has single critical point but the
gap vanishes at a surface in the BZ. The above two features are expected
in high dimensional or multiple parameter-dependent systems, such
as the nodal line semimetal$\;$\citep{Xie2015APLM,Chan2016prb,Wang2017prb}.
An trivial example of the first case is the 1D Kitaev chain$\;$(or
1D XY model)$\;$\citep{Divakaran2009jsm,Deng2009prb}, which has
gapless phase when the superconducting gap is zero. When quenching
across this gapless phase, the gapless point sweeps the 1D BZ, forming
a 1D critical surface. The work and quasiparticle excitation are all
constant, consistent with the prediction of Eq.$\;$(\ref{eq:scal_work_crisur}).
(3)$\;$We consider only the situation of quenching the system from
one gapped phase to another. Other protocols can also be considered
such as stopping at an anisotropic QCP$\;$\citep{Hikichi2010prb}
or periodically driven protocols$\;$\citep{Dutta2015pre}. Hopefully,
different scaling behaviors of cumulants of work will be obtained
for these protocols.

In summary, we have extended the scaling behavior for work statistics
form crossing an isolated QCP to a critical surface. The presence
of the critical surface reduces the available phase space for quasiparticle
excitation, resulting in a scaling behavior for work statistics different
from the situation of an isolated QCP. We use the adiabatic perturbation
theory to study the general scaling behavior and calculate the exact
CFW of the KHM by utilizing the group-theoretical technique to verify
our general results. The exact CFW shows that the work distribution
is a Poisson binomial distribution. Extensions of our current work
to systems beyond the quasiparticle picture and single-particle excitation
approximation will be given in our future studies.
\begin{acknowledgments}
Fan Zhang thanks the helpful discussion with Jinfu Chen and Zhaoyu
Fei. H. T. Quan acknowledges support from the National Science Foundation
of China under grants 11775001, 11534002, and 11825501.
\end{acknowledgments}

\appendix

\section{Landau-Zener Problem of KHM\label{sec:Landau-Zener-Problem-of}}

In this appendix, we sketch the solution to Eq.$\;$(\ref{eq:L and C}).
Since $\tau_{F}\mcH_{k}$ in Eq.$\;$(\ref{eq:L and C}) is block
diagonal, $L_{k}(t)$ and $C_{k}(t)$ will remain block diagonal in
the evolution as they are initially block diagonal. We do not need
to consider $C_{k}(t)$, because $C_{k}(t)$ is a zero matrix initially
and will remain zero under the unitary evolution. It is sufficient
to consider the evolution of the upper block matrix $\Lambda_{k}(t)$
of $L_{k}(t)$. The lower block matrix is related to the upper block
matrix by the complex conjugate. The upper block matrix of $\tau_{F}\mcH_{k}$
is 
\begin{equation}
\mcH_{k}^{u}=\begin{pmatrix}0 & -if\\
if^{*} & 0
\end{pmatrix}_{k}=\epsilon_{k}\sigma^{y}+\Delta_{k}\sigma^{x}.\label{eq:Ham_upper}
\end{equation}
The dynamics can be transformed to the standard form of the Landau-Zener
problem by a gauge transformation $U_{x}=\exp(-i\pi\sigma^{x}/4)$
and introducing a rescaled time $s_{k}=\sqrt{2v}t+\sqrt{2}(J_{x}\cos k_{1}+J_{y}\cos k_{2})/v$$\;$\citep{Sengupta2008prl,Mondal2008prb,Hikichi2010prb}.
For simplicity, we omit the subscript $k$ of $s_{k}$. The equation
of motion of $\tilde{\Lambda}_{k}(s)=U_{x}\Lambda_{k}(t)$ is 
\[
i\frac{d}{ds}\tilde{\Lambda}_{k}(s)=\begin{pmatrix}s & \sqrt{2}\Delta_{k}/\sqrt{v}\\
\sqrt{2}\Delta_{k}/\sqrt{v} & s
\end{pmatrix}\Lambda_{k}'.
\]
 In the limit $t_{0}\to-\infty,t_{1}\to\infty$, i.e., $s_{0}\equiv s(t_{0})\to-\infty,s_{1}\equiv s(t_{1})\to\infty,$
the asymptotic solution of $\tilde{\Lambda}_{k}(s)$ is
\begin{widetext}
\[
\tilde{\Lambda}_{k}(s_{1})=\bigg\{\sqrt{p_{k}}\left[\cos\tilde{\Theta}_{k}(s_{1},s_{0})\bfI_{2\times2}+i\sin\tilde{\Theta}_{k}(s_{1},s_{0})\sigma^{z}\right]+i\sqrt{1-p_{k}}\left[\cos\tilde{\Theta'}_{k}(s_{1},s_{0})\sigma^{y}+\sin\tilde{\Theta'}_{k}(s_{1},s_{0})\sigma^{x}\right]\bigg\}\tilde{\Lambda}_{k}(s_{0}),
\]
 where $\tilde{\Lambda}_{k}(s_{0})=U_{x}$, Landau-Zener transition
probability $p_{k}=\exp(-2\pi\Delta_{k}^{2}/v)$ and 
\begin{align}
\tilde{\Theta}(s_{1},s_{0}) & =-\frac{i}{2}(s_{1}^{2}-s_{0}^{2})+\frac{i\Delta_{k}^{2}}{v}\ln(-s_{0}/s_{1}),\nonumber \\
\tilde{\Theta}'(s_{1},s_{0}) & =i\pi/4-\frac{i}{2}(s_{1}^{2}+s_{0}^{2})-\frac{i\Delta_{k}^{2}}{v}\ln(-2s_{0}s_{1})+\ln\left[\Gamma(i\Delta_{k}^{2}/v)\sqrt{\Delta_{k}^{2}\sinh(\pi\Delta_{k}^{2}/v)/(v\pi)}\right]\label{eq:theta_app1}
\end{align}
with gamma function $\Gamma(x)$. Changing the variable from $s$
to $t$ in Eq.$\;$(\ref{eq:theta_app1}), we obtain the expression
of $\Theta(t_{1},t_{0})$ and $\Theta'(t_{1},t_{0})$ in Eq.$\;$(\ref{eq:lam_upp}).
\end{widetext}

\section{Quench along one edge of the gapless phase\label{sec:Quench-through-one-edge}}

In this appendix, we consider the protocol that quenches along one
edge of the shaded triangle$\;${[}see Fig.$\;$\ref{fig:The-phase-diagram}$\;$(III){]},
i.e. we keep $J_{x}=J_{y}+J_{z}$ and linearly vary $J_{z}=vt$ and
$J_{y}=J_{x}-vt.$ The system is always gapless along the edge. The
Hamiltonian of the corresponding Landau-Zener problem is 
\[
\hatH_{LZ}(\bfk)=2\lrb{vt\sin\frac{k_{2}}{2}+\tilde{\epsilon}(\bfk)}\hat{\sigma}_{z}+2\tilde{\Delta}(\bfk)\hat{\sigma}_{\perp}(\bfk),
\]
 where $\tilde{\epsilon}(\bfk)=2J_{x}\sin[(2k_{2}-k_{1})/2]\cos[(k_{1}+k_{2})]/2,$
and the minimum gap $2\tilde{\Delta}(\bfk)=2J_{x}\cos[(k_{1}-2k_{2})/2]\cos\left[\left(k_{1}+k_{2}\right)/2\right].$
The matrix $\hat{\sigma}_{\perp}(\bfk)=-\sin\lrp{k_{2}/2}\hat{\sigma}_{x}+\cos\lrp{k_{2}/2}\hat{\sigma}_{y}.$
The Landau-Zener probability for this protocol is 
\begin{align*}
p_{k}' & =\exp\left[-\frac{2\pi\tilde{\Delta}^{2}(\bfk)}{v\sin(k_{2}/2)}\right]\\
 & =\exp\left[-\frac{8\pi J_{x}^{2}\cos^{2}\left(\frac{k_{1}-2k_{2}}{2}\right)\cos^{2}\left(\frac{k_{1}+k_{2}}{2}\right)}{v\sin(k_{2}/2)}\right].
\end{align*}
Corresponding to this protocol, there are three critical lines in
which the minimum gap $\tilde{\Delta}$ vanishes: $k_{x}=\pi/\sqrt{3}$,
$k_{x}=3\sqrt{3}k_{y}\pm2\pi/\sqrt{3}.$ Expand $\tilde{\Delta}(\bfk)$
along these critical lines, we get $\tilde{\Delta}(\bfk)\sim|\bfk|\equiv|\bfk|^{z_{2}}$
and $\tilde{\epsilon}(\bfk)\sim|\bfk|^{2}\equiv|\bfk|^{z_{1}}$. Hence,
the dynamical critical exponent is still $z=1$. Since this protocol
also draws critical lines in the HBZ, we conclude that its scaling
behavior of work statistics is the same as quenching across the inner
area of the gapless phase, i.e., $\kappa_{n}\sim v^{1/2}.$ This is
different from the case in 1D XY model when quenching along the gapless
line$\;$\citep{Divakaran2009jsm,Deng2009prb,Mondal2009prb}, where
$z=z_{1}=1<z_{2}=2$. In the latter case, the mean density of defects
scales as $\avg{n_{ex}}\sim v^{d\nu/(\nu z_{2}+1)}=v^{1/3}$ rather
than $v^{d\nu/(z\nu+1)}=v^{1/2}$ predicted by KZM.

\bibliographystyle{apsrev4-1}
\bibliography{ref}

\end{document}